\documentclass[journal,twocolumn,10pt]{IEEEtran}
\usepackage{amsfonts}
\usepackage{amsfonts}
\usepackage{amsfonts}
\usepackage{amsfonts}
\usepackage[latin1]{inputenc}
\usepackage{amssymb}
\usepackage[usenames]{color}
\usepackage{latexsym}
\usepackage{mathrsfs}
\usepackage{amsfonts}
\usepackage{amsbsy}
\usepackage{amsmath}
\usepackage{times}
\usepackage{epsfig,epsf,times,amssymb,amsmath}

\begin{document}
\title{Direction Finding Algorithms with Joint Iterative Subspace Optimization }

\author{Lei~Wang, Rodrigo~C.~de~Lamare, and
        Martin~Haardt
\thanks{Part of this work has been presented a the IEEE International Conference on Acoustics, Speech,
and Signal Processing,
 2010, pp. 2626-2629, 2010 \cite{Wang}. Lei Wang and Rodrigo C. de Lamare are with Department
of Electronics, The University of York, York, YO10 5DD, U.K.
(e-mail: lw517@york.ac.uk, rcdl500@ohm.york.ac.uk).}\\
\thanks{Martin Haardt is with Communication Research Laboratory, Ilmenau University of Technology,
D-98684, Germany (e-mail: martin.haardt@tu-ilmenau.de).}}

\maketitle

\begin{abstract}

In this paper, a reduced-rank scheme with joint iterative
optimization is presented for direction of arrival estimation. A
rank-reduction matrix and an auxiliary reduced-rank parameter vector
are jointly optimized to calculate the output power with respect to
each scanning angle. Subspace algorithms to estimate the
rank-reduction matrix and the auxiliary vector are proposed.
Simulations are performed to show that the proposed algorithms
achieve an enhanced performance over existing algorithms in the
studied scenarios.
\end{abstract}


\section{Introduction}
In many array processing related fields such as radar, sonar, and
wireless communications, the information of interest extracted from
the received signals is the direction of arrival (DOA) of waves
transmitted from radiating sources to the antenna array. The DOA
estimation problem has received considerable attention in the last
several decades \cite{Krim}. Many estimation algorithms have been
reported in the literature, e.g., \cite{Liberti},
 {\cite[Chapters 8 and 9]{Trees}}, and the references
therein. Among the most representative algorithms are Capon's method
\cite{Capon}, maximum-likelihood (ML) \cite{Ziskind}, and
subspace-based schemes\cite{Schmidt}-\cite{Haardt2}.

 {Capon's method calculates the output power spectrum
over the scanning angles and determines the DOA by locating the
peaks in the spectrum. The implementation is relatively simple. The
drawback of this method is that the resolution strongly depends on
the number of available snapshots, the signal-to-noise ratio (SNR)
and the array size. The ML type algorithms are robust for DOA
estimation since they exhibit superior resolution in hostile
scenarios with a low input SNR as long as the number of snapshots is
small. Moreover, they work well when the sources are correlated.
However, the implementation of the ML type methods is complicated
and requires intensive computational cost, which limits their
practical applications.}

\subsection{Prior Work}

 {Subspace-based algorithms, which exploit the
structure of the received data to decompose the observation space
into a signal subspace and a corresponding orthogonal noise
subspace, play an important role for DOA estimation. According to
the approach to compute the signal subspace, the subspace-based
methods can be classified into eigen-decomposition, subspace
tracking, and basis vectors based algorithms. Among the most popular
and cost-effective eigen-decomposition algorithms are MUSIC
\cite{Schmidt} and ESPRIT \cite{Roy} that require an
eigen-decomposition. The MUSIC algorithm computes the output power
spectrum by scanning the possible angles and selects the peaks to
estimate the directions of the sources. The root MUSIC algorithm
 { \cite[pp. 1158-1163]{Trees}} and its low-complexity
versions \cite{ren} have also been reported and shown to result in
efficient DOA estimates. The ESPRIT algorithm employs a displacement
invariance in some specific array structures and requires a lower
complexity than MUSIC  { \cite[pp. 1170-1194]{Trees}.} Iterative DOA
estimation methods exploiting the removal of detected signals have
been reported in the literature \cite{morrison}, whereas adaptive
techniques based on the multistage Wiener filter and that operates
in the Krylov subspace have been considered in
\cite{tureli,delamare11}. Subspace tracking techniques (e.g.,
approximated power iteration (API)) \cite{Yang}-\cite{Badeau} avoid
a direct eigen-decomposition and employ an iterative procedure to
estimate the signal subspace. These techniques can effectively
reduce the computational complexity but often result in some
performance degradation. Another recent class of subspace algorithms
include those that employ basis vectors such as the auxiliary vector
(AV) \cite{Grover}, the conjugate gradient (CG)
\cite{Semira,steinwandt} and iterative procedures
\cite{tvt2010}-\cite{jidf}, which construct the signal subspace
using an iterative procedure without resorting to an
eigen-decomposition.}

\subsection{Contributions}

 {In this paper, a novel reduced-rank scheme is
presented and adaptive algorithms for DOA estimation are developed
for scenarios with a small number of snapshots.} The reduced-rank
scheme consists of a rank-reduction matrix, which is responsible for
mapping the received vector into a lower dimension, and an auxiliary
reduced-rank parameter vector that is employed to calculate the
output power with respect to each scanning angle. Unlike previous
techniques that either require an eigen-decomposition, the use of a
subspace tracking algorithms, or an iterative procedure to compute
the basis vectors, the proposed method computes the rank-reduction
matrix and the auxiliary reduced-rank parameter vector based on a
least-squares optimization algorithm along with an alternating
procedure between the rank-reduction step and the computation of the
auxiliary reduced-rank parameter vector. The rank-reduction matrix
and the auxiliary reduced-rank parameter vector are jointly
optimized according to the minimum variance (MV) design criterion
for computing the output power spectrum.   {The polynomial rooting
technique \cite{Tyrtyshnikov} is employed in the proposed scheme to
estimate the DOAs without an exhaustive search through all possible
angles. }  { We derive a constrained least squares (LS) based
algorithm to iteratively estimate the rank-reduction matrix and the
auxiliary reduced-rank parameter vector. The proposed algorithm,
which is termed joint iterative optimization (JIO), provides an
iterative exchange of information between the rank reduction matrix
and the reduced-rank vector and thus leads to an improved resolution
\cite{Wang}. The complexity of the proposed JIO algorithm can be
reduced without any significant degradation of the resolution by
utilizing the matrix inversion lemma \cite{Haykin} or resorting to
optimization algorithms with lower computation cost, i.e.,
stochastic gradient techniques. Other approaches based on the QR
decomposition are also possible for implementation
 {\cite[pp. 779]{Trees}.} A model-order selection
approach is developed to select the most adequate rank for the
proposed JIO algorithms to ensure the best performance is obtained.
A version of the proposed JIO algorithms with forward/backward
averaging (FBA) \cite{Evans}, \cite{Pillai} is also devised to deal
with highly correlated sources. The proposed algorithms are suitable
for DOA estimation with large arrays, dynamic scenarios in which the
DoA changes over time and a small number of snapshots, and exhibit
an advantage over existing algorithms in the presence of many
sources. We conduct a study that shows that Capon's and
subspace-based methods are inferior to the proposed JIO algorithms
for a sufficiently large array. Although the ML algorithm is robust
to these conditions, with large arrays it has an extremely high
computational cost which prevents its use practice. Furthermore, the
proposed JIO algorithms work well without an exact knowledge of the
number of sources, which significantly degrades the performance of
the subspace-based and the ML methods. }

In summary, this paper makes the following contributions:
\begin{itemize}
\item A reduced-rank scheme is introduced for DOA estimation. A joint optimization strategy between
the rank-reduction matrix and the auxiliary reduced-rank parameter
vector based on the MV criterion is employed for improving the
resolution.
\item Reduced-rank DOA estimation algorithms are proposed. The FBA technique is
applied to the proposed JIO algorithms to deal with correlated
sources.
\item We develop a model-order selection approach to select the best
rank for the proposed JIO algorithms. A comparison is presented to
show the computational complexity of the proposed and existing DOA
estimation algorithms.
\item A simulation study is performed to show the improved resolution of
the proposed JIO algorithms over existing ones in a number of
scenarios of practical interest.
\end{itemize}

This paper is structured as follows: we outline a system model for
DOA estimation in Section II. The proposed reduced-rank scheme and
the application of the polynomial rooting technique are introduced
in Section III. In Section IV, we derive the proposed JIO algorithms
and illustrate the use of the model-order selection and the FBA
techniques. A complexity analysis is also presented in this section.
Simulation results are provided and discussed in Section V, and
conclusions are drawn in Section VI.

\section{System Model}

 {Let us suppose that $q$ narrowband signals impinge
on a uniform linear array (ULA) of $M$ ($M\geq q$) sensor elements.
It should be remarked that the proposed DOA estimation algorithm can
be applied to arbitrary array structures. The ULA is adopted here
for using the FBA and polynomial rooting techniques and providing a
fair comparison with ESPRIT, which has been developed for some
specific array structures. The $i$th received vector of the array
output $\boldsymbol x(i)\in\mathbb C^{M\times 1}$ can be modeled as}
\begin{equation} \label{1}
\centering {\boldsymbol x}(i)={\boldsymbol A}({\boldsymbol
{\theta}}){\boldsymbol s}(i)+{\boldsymbol n}(i),~~~ i=1,\ldots,N,
\end{equation}
where
$\boldsymbol{\theta}=[\theta_{1},\ldots,\theta_{q}]^{T}\in\mathbb{C}^{q
\times 1}$ contains the DOAs of the signals, ${\boldsymbol
A}({\boldsymbol {\theta}})=[{\boldsymbol
a}(\theta_{1}),\ldots,{\boldsymbol a}(\theta_{q})]\in\mathbb{C}^{M
\times q}$ is the matrix that contains the steering vectors
${\boldsymbol a}(\theta_{k})$, where ${\boldsymbol
a}(\theta_{k})=[1,e^{-2\pi
j\frac{d}{\lambda_{c}}cos{\theta_{k}}},\ldots, e^{-2\pi
j(M-1)\frac{d}{\lambda_{c}}cos{\theta_{k}}}]^{T}\in\mathbb{C}^{M
\times 1},~~~(k=1,\ldots,q)$, $\lambda_{c}$ is the wavelength, $d$
($d=\lambda_{c}/2$ in general) is the inter-element distance of the
ULA, ${\boldsymbol s}(i)\in \mathbb{R}^{q\times 1}$ contains the
source symbols, ${\boldsymbol n}(i)\in\mathbb{C}^{M\times 1}$ is the
white sensor noise that is assumed to be a zero-mean spatially
uncorrelated and Gaussian process, $N$ is the number of snapshots,
and $(\cdot)^{T}$\ denotes transpose. To avoid mathematical
ambiguities, the steering vectors $\boldsymbol a(\theta_{k})$ are
considered to be linearly independent
 {\cite[pp.845]{Trees}.}

The spatial correlation matrix of the received vector is
\begin{equation}\label{2}
\boldsymbol R=\mathbb E[\boldsymbol x(i)\boldsymbol
x^{H}(i)]=\boldsymbol A(\boldsymbol \theta)\boldsymbol
R_{s}\boldsymbol {A}^{H}(\boldsymbol
\theta)+\sigma_{n}^{2}\boldsymbol I,
\end{equation}
where $\boldsymbol R_{s}=\mathbb E[\boldsymbol s(i)\boldsymbol
s^{H}(i)]$ denotes the signal covariance matrix, which is diagonal
if the sources are uncorrelated and is nondiagonal and nonsingular
for partially correlated sources, $\mathbb E[\boldsymbol
n(i)\boldsymbol n^{H}(i)]=\sigma_{n}^{2}\boldsymbol I_{M\times M}$
with $\boldsymbol I_{M\times M}$ being the corresponding identity
matrix, and $(\cdot)H$ denotes Hermitian transpose.
 {It is well understood in the literature  { \cite[pp.
 1204]{Trees} }
that a small number of snapshots results in a poor estimate of the
correlation matrix, which degrades the DOA estimation resolution of
Capon's method and most subspace-based methods. With large arrays,
the resolution can be compensated to a certain extent whereas the
computational cost increases. Moreover, the performance of
eigen-decomposition and subspace tracking based methods is affected
by highly correlated sources. In these situations, the use of the
FBA technique can mitigate the performance degradation caused by a
high level of correlation between the sources. The recent AV and CG
algorithms can also deal with the problem of correlated sources but
lose their superiority when a large number of sources need to be
located.}

\section{Proposed Reduced-Rank Scheme}

In this section, we introduce a reduced-rank strategy with the MV
criterion to obtain the output power spectrum with respect to the
possible scanning angles and find the peaks for DOA estimation. The
polynomial rooting technique is employed in the new scheme to
circumvent an exhaustive search that leads to a reduced
computational complexity.

\subsection{Proposed Reduced-rank Scheme for DOA estimation}

\begin{figure}[h]
    \centerline{\psfig{figure=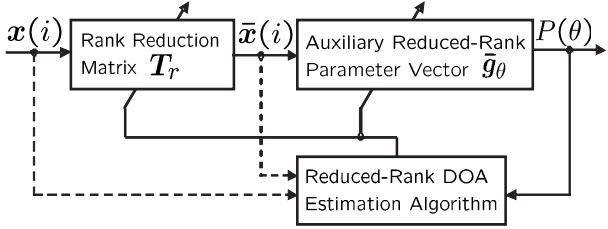,height=30mm,width=80mm} }
    \caption{Proposed reduced-rank structure.}
    \label{fig:DOA}
\end{figure}

The proposed reduced-rank structure is depicted in Fig.
\ref{fig:DOA}. We introduce a rank-reduction matrix $\boldsymbol
T_r\in\mathbb C^{M\times r}$, which maps the full-rank received
vector $\boldsymbol x(i)$ into a lower dimension and generates the
reduced-rank received vector $\bar{\boldsymbol x}(i)\in\mathbb
C^{r\times1}$
\begin{equation}\label{4}
\bar{\boldsymbol x}(i)=\boldsymbol T_r^H\boldsymbol x(i),
\end{equation}
where $\boldsymbol T_r$ consists of a collection of $r$
$M$-dimensional vectors $\boldsymbol t_l=[t_{1,l}, t_{2,l}, \ldots,
t_{M,l}]^T\in\mathbb C^{M\times1}$, $l=1,\ldots,r$ as given by
$\boldsymbol T_r=[\boldsymbol t_1, \boldsymbol t_2, \ldots,
\boldsymbol t_r]$, and $r$ is the rank that is assumed to be less
than $M$. In what follows, all $r$ dimensional quantities are
denoted with a ``bar". Compared with $\boldsymbol x(i)$, the
dimension of $\bar{\boldsymbol x}(i)$ is reduced and the key
features of the original signal are retained in $\bar{\boldsymbol
x}(i)$ according to the design criterion. An auxiliary filter with
the reduced-rank vector $\bar{\boldsymbol g}_{\theta}=[\bar{
g}_{\theta,1}, \bar{g}_{\theta,2}, \ldots, \bar{
g}_{\theta,r}]^T\in\mathbb C^{r\times1}$ is used after the rank
reduction matrix to process $\bar{\boldsymbol x}(i)$ to compute the
output power with respect to the current scanning angle. The
computational complexity is reduced if $r << M$ for large arrays.

 {The rank-reduction matrix $\boldsymbol T_r$ and the
auxiliary reduced-rank parameter vector $\bar{\boldsymbol
g}_{\theta}$ are computed by the following optimization problem}
\begin{equation}\label{5}
\begin{split}
&\hat{\theta}=\arg\min_{\bar{\boldsymbol g}_{\theta},~\boldsymbol
T_r }~~\bar{\boldsymbol g}_{\theta}^H\boldsymbol T_r^H {\boldsymbol
R}\boldsymbol
T_r\bar{\boldsymbol g}_{\theta}\\
&\textrm{subject~to}~~\bar{\boldsymbol g}_{\theta}^H\boldsymbol
T_r^H\boldsymbol a(\theta)=1,
\end{split}
\end{equation}
where ${\boldsymbol R}$ is the covariance matrix and the
optimization problem depends on $\boldsymbol T_r$ and
$\bar{\boldsymbol g}_{\theta}$, which have to be estimated with
respect to $\theta$.

The optimization problem in (\ref{5}) can be transformed by the
method of Lagrange multiplier into an unconstrained one, which is
\begin{equation}\label{6}
\mathcal J=\bar{\boldsymbol g}_{\theta}^{H}{\boldsymbol T}_r^H
{\boldsymbol R} {\boldsymbol T}_r \bar{\boldsymbol g}_{\theta}+2~
\mathfrak{R}~\big\{\lambda[\bar{\boldsymbol g}_{\theta}^H\boldsymbol
T_r^H\boldsymbol a(\theta)-1]\big\},
\end{equation}
where $\lambda$ is a scalar Lagrange multiplier and the operator
$\mathfrak{R}(\cdot)$ selects the real part of the argument.

 {In order to obtain $\boldsymbol T_r$ and
$\bar{\boldsymbol g}_{\theta}$, we make an assumption that one
quantity is known and compute the other one. Specifically, assuming
$\bar{\boldsymbol g}_{\theta}$ is known and taking the gradient of
(\ref{6}) with respect to $\boldsymbol T_r^{\ast}$, where $\ast$
denotes complex conjugate, we have}
\begin{equation}\label{7}
\nabla\mathcal J_{T_r^{\ast}}=\boldsymbol R\boldsymbol
T_r\bar{\boldsymbol g}_{\theta}\bar{\boldsymbol
g}_{\theta}^H+\lambda_{T_r^{\ast}}\boldsymbol
a(\theta)\bar{\boldsymbol g}_{\theta}^H.
\end{equation}
Equating the gradient to a zero matrix and solving for
$\lambda_{T_r^{\ast}}$, the rank-reduction matrix can be expressed
as
\begin{equation}\label{8}
\boldsymbol T_r=\frac{\boldsymbol R^{-1}\boldsymbol
a(\theta)}{\boldsymbol a^H(\theta)\boldsymbol R^{-1}\boldsymbol
a(\theta)}\frac{\bar{\boldsymbol g}_{\theta}^H}{\|\bar{\boldsymbol
g}_{\theta}\|^2},
\end{equation}
where for a small number of snapshots, $\boldsymbol R^{-1}$ is
estimated by either employing diagonal loading or the
pseudo-inverse. The derivation of (\ref{8}) is given in the
Appendix.

 {Assuming that $\boldsymbol T_r$ is known and taking
the gradient of (\ref{6}) with respect to $\bar{\boldsymbol
g}_{\theta}^{\ast}$, we have
\begin{equation}\label{9}
\begin{split}
\nabla\mathcal J_{\bar{g}_{\theta}^{\ast}} & = {\boldsymbol T}_r^H
{\boldsymbol R} {\boldsymbol T}_r \bar{\boldsymbol g}_{\theta} +
\lambda_{\bar{g}_{\theta}^{\ast}}\boldsymbol T_r^H\boldsymbol
a(\theta)\\ & = \bar{\boldsymbol R}\bar{\boldsymbol g}_{\theta} +
\lambda_{\bar{g}_{\theta}^{\ast}}\bar{\boldsymbol a}(\theta),
\end{split}
\end{equation}
where $\bar{\boldsymbol R}=\mathbb E[\bar{\boldsymbol
x}(i)\bar{\boldsymbol x}^H(i)]\in\mathbb C^{r\times r}$ is the
reduced-rank covariance matrix.}  {Setting \eqref{9} equal to a zero
vector and solving for $\lambda_{\bar{g}_{\theta}^{\ast}}$, we
obtain
\begin{equation}\label{10}
\bar{\boldsymbol g}_{\theta}=\frac{\bar{\boldsymbol
R}^{-1}\bar{\boldsymbol a}(\theta)}{\bar{\boldsymbol
a}^H(\theta)\bar{\boldsymbol R}^{-1}\bar{\boldsymbol a}(\theta)},
\end{equation}
where $\bar{\boldsymbol a}(\theta)=\boldsymbol T_r^H\boldsymbol
a(\theta)\in\mathbb C^{r\times1}$ is the reduced-rank steering
vector with respect to the current scanning angle. A detailed
derivation is included in the Appendix. Note that the auxiliary
reduced-rank vector $\bar{\boldsymbol g}_{\theta}$ is more general
for dealing with DOA estimation. Specifically, for $r=1$, the
proposed JIO algorithm results in Capon's method. For $1<r<M$, it
operates under a lower dimension and thus reduces the complexity.}

The output power for each scanning angle is calculated by
substituting the expressions of $\boldsymbol T_r$ in (\ref{8}) and
$\bar{\boldsymbol g}_{\theta}$ in (\ref{10}) into (\ref{5}), which
yields
\begin{equation}\label{11}
P(\theta)=\big(\bar{\boldsymbol a}^H(\theta)\bar{\boldsymbol
R}^{-1}\bar{\boldsymbol a}(\theta)\big)^{-1}.
\end{equation}

By searching all possible angles, we could find peaks in the output
power spectrum that correspond to the DOAs of the sources.

\subsection{Proposed Scheme with Polynomial Rooting}

 {In order to avoid the exhaustive search through all
possible angles, we take the polynomial rooting technique into
account. Specifically, premultiplying the terms in (\ref{10}) by
$\bar{\boldsymbol g}_{\theta}^H$ and rearranging the terms, we have
\begin{equation}\label{12}
Q(\theta)=\bar{\boldsymbol a}^H(\theta)\bar{\boldsymbol
R}^{-1}\bar{\boldsymbol a}(\theta)=\frac{\bar{\boldsymbol
g}_{\theta}^H\bar{\boldsymbol R}^{-1}\bar{\boldsymbol
a}(\theta)}{\|\bar{\boldsymbol g}_{\theta}\|^2},
\end{equation}
where $Q(\theta)=P^{-1}(\theta)$.}

The proposed reduced-rank scheme performs DOA estimation by scanning
a limited range of angles without an exhaustive search. Compared
with (\ref{11}), the expression in (\ref{12}) brings a
simplification in the joint optimization of the rank-reduction
matrix and the auxiliary reduced-rank parameter vector. We show the
advantage of this scheme in the simulations. Note that, in this
paper, the objective of the application of the polynomial rooting is
not only an extension of the proposed scheme. It is viewed as an
approach to reduce the computational complexity.

Another effective technique that could be employed in the proposed
scheme is beamspace preprocessing  {\cite[pp. 1243-1251]{Trees}},
\cite{Gansman}, which preprocesses the received vector with a matrix
that, in essence, creates a set of beams. It reduces the complexity
from the number of sensor elements to the number of beams utilized
to probe a given sector. We then use these beam outputs to estimate
the DOAs. In many environments, DOA estimation with the beamspace
preprocessing achieves an improved probability of resolution with
substantially less computational complexity. Furthermore, numerous
works have been reported to combine polynomial rooting and beamspace
techniques into one scheme for robustness \cite{Zoltowski},
\cite{Gershman}. In the proposed reduced-rank scheme, it is possible
to consider the beamspace technique as an extension of the current
work for improving the resolution and reducing the complexity.

\section{Proposed Reduced-Rank Algorithms}

In this section, we derive a constrained LS algorithm for an
implementation of the proposed reduced-rank scheme. The proposed
algorithm jointly estimates the rank-reduction matrix and auxiliary
reduced-rank parameter vector using an alternating optimization
procedure. The rank is selected via the model order selection
approach. The FBA technique is employed in the algorithm to deal
with highly correlated sources for the resolution improvement. We
utilize the matrix inversion lemma to develop a recursive least
squares (RLS) based algorithm for DOA estimation.
 {With these algorithms a designer can choose between
batch or adaptive (recursive) processing. In particular, adaptive
techniques can be used if a designer is interested in reducing the
computational cost per snapshot as compared to computing a matrix
inversion. In batch processing a designer needs to compute a matrix
inversion, which might be a choice for a stationary scenario that
only requires one matrix inversion.}

\subsection{Proposed JIO Algorithm}
From (\ref{8}) and (\ref{10}), the challenge left to us is how to
efficiently compute the rank-reduction matrix $\boldsymbol T_r$ and
the auxiliary reduced-rank vector $\bar{\boldsymbol g}_{\theta}$ for
solving (\ref{5}). Using the method of LS, the constraint in
(\ref{5}) can be incorporated by the method of Lagrange multipliers
in the form {\small
\begin{equation}\label{17}
\mathcal J_{\textrm{LS}}=\sum_{l=1}^{i}\alpha^{i-l}|\bar{\boldsymbol
g}_{\theta}^H(i)\boldsymbol T_r^H(i)\boldsymbol
x(l)|^2+2~\mathfrak{R}\big\{\lambda\big[\bar{\boldsymbol
g}_{\theta}^H(i)\boldsymbol T_r^H(i)\boldsymbol
a(\theta)-1\big]\big\},
\end{equation}}
where $\alpha$ is a forgetting factor that is a positive constant
close to, but less than $1$. Assuming $\bar{\boldsymbol
g}_{\theta}(i)$ is known, taking the gradient of (\ref{17}) with
respect to $\boldsymbol T_r^{\ast}(i)$ yields,
\begin{equation}\label{18}
\begin{split}
{\nabla\mathcal
J_{\textrm{LS}_{T_r^{\ast}}}}&=\sum_{l=1}^{i}\alpha^{i-l}\boldsymbol
x(l)\boldsymbol x^H(l)\boldsymbol T_r(i)\bar{\boldsymbol
g}_{\theta}(i)\bar{\boldsymbol
g}_{\theta}^H(i)+\lambda_{T_r^{\ast}}\boldsymbol
a(\theta)\bar{\boldsymbol
g}_{\theta}^H(i)\\
&=\hat{\boldsymbol R}(i)\boldsymbol T_r(i)\bar{\boldsymbol
g}_{\theta}(i)\bar{\boldsymbol
g}^H(\theta)(i)+\lambda_{T_r^{\ast}}\boldsymbol
a(\theta)\bar{\boldsymbol g}_{\theta}^H(i),
\end{split}
\end{equation}
where $\hat{\boldsymbol R}(i)=\sum_{l=1}^{i}\alpha^{i-l}\boldsymbol
x(l)\boldsymbol x^H(l)\in\mathbb C^{M\times M}$ is an estimate of
the covariance matrix at time instant $i$ and can be written in a
recursive form $\hat{\boldsymbol R}(i)=\alpha\hat{\boldsymbol
R}(i-1)+\boldsymbol x(i)\boldsymbol x^H(i)$.

The resulting rank-reduction matrix is
\begin{equation}\label{19}
\boldsymbol T_r(i)=\frac{\hat{\boldsymbol R}^{-1}(i)\boldsymbol
a(\theta)}{\boldsymbol a^H(\theta)\hat{\boldsymbol
R}^{-1}(i)\boldsymbol a(\theta)}\frac{\bar{\boldsymbol
g}_{\theta}^H(i)}{\|\bar{\boldsymbol g}_{\theta}(i)\|^2}.
\end{equation}

Fixing $\boldsymbol T_r(i)$, taking the gradient of (\ref{17}) with
respect to $\bar{\boldsymbol g}_{\theta}^{\ast}(i)$, it becomes
{\small
\begin{equation}\label{20}
\begin{split}
\nabla\mathcal
J_{\textrm{LS}_{\bar{g}_{\theta}^{\ast}}}&=\sum_{l=1}^{i}\alpha^{i-l}\boldsymbol
T_r^H(i)\boldsymbol x(l)\boldsymbol x^H(l)\boldsymbol
T_r(i)\bar{\boldsymbol
g}_{\theta}(i)+\lambda_{\bar{g}_{\theta}^{\ast}}\boldsymbol
T_r^H(i)\boldsymbol a(\theta)\\
&=\hat{\bar{\boldsymbol R}}(i)\bar{\boldsymbol
g}_{\theta}(i)+\lambda_{\bar{g}_{\theta}^{\ast}}\boldsymbol
T_r^H(i)\boldsymbol a(\theta),
\end{split}
\end{equation}}
where $\hat{\bar{\boldsymbol
R}}(i)=\sum_{l=1}^{i}\alpha^{i-l}\bar{\boldsymbol
x}(l)\bar{\boldsymbol x}^H(l)\in\mathbb C^{r\times r}$ is an
estimate of the reduced-rank covariance matrix. Its recursive form
is $\hat{\bar{\boldsymbol R}}(i)=\alpha\hat{\bar{\boldsymbol
R}}(i-1)+\bar{\boldsymbol x}(i)\bar{\boldsymbol x}^H(i)$. The
resulting expression of $\bar{\boldsymbol g}_{\theta}(i)$ is
\begin{equation}\label{21}
\bar{\boldsymbol g}_{\theta}(i)=\frac{\hat{\bar{\boldsymbol
R}}^{-1}(i)\bar{\boldsymbol a}(\theta)}{\bar{\boldsymbol
a}^H(\theta)\hat{\bar{\boldsymbol R}}^{-1}(i)\bar{\boldsymbol
a}(\theta)}.
\end{equation}
 {Note that the expression of the rank-reduction
matrix in (\ref{19}) is a function of $\bar{\boldsymbol
g}_{\theta}(i)$ while the auxiliary reduced-rank vector obtained
from (\ref{21}) depends on $\boldsymbol T_r(i)$. The proposed
algorithm relies on an iterative exchange of information between
$\boldsymbol T_r(i)$ and $\bar{\boldsymbol g}_{\theta}(i)$, which
results in an improved convergence performance. The proposed JIO
algorithm is summarized in Table \ref{tab:JIO}, where
$\hat{\bar{\boldsymbol R}}$ is the estimate of the reduced-rank
correlation matrix related to $N$ snapshots, the scanning angle
$\theta_n=n\triangle^o$, $\triangle^o$ is the search step, and
$n=1,2, \ldots, 180^o/\triangle^o$. For a simple and convenient
search, we make $180^o/\triangle^o$ an integer. It is necessary to
initialize $\boldsymbol T_r(0)$ to start the update due to the
dependence between $\boldsymbol T_r(i)$ and $\bar{\boldsymbol
g}_{\theta}(i)$, see Table \ref{tab:JIO}.}

The output power $P(\theta_n)$ is much higher if the scanning angle
$\theta_n=\theta_k$, ($k=0, \ldots, q-1$), which corresponds to the
position of the source, compared with other scanning angles with
respect to the noise level. Thus, we can estimate the DOAs by
finding the peaks in the output power spectrum. We refer to
 { \cite[pp.1142-1146]{Trees} } for the individual
computational costs of the recursions.

\begin{table}[t]
\centering
    \caption{The proposed JIO algorithm}
    \label{tab:JIO}
    \begin{small}
    \begin{tabular}{|l|}
\hline
\bfseries {Initialization:}\\
~~~~~~$\boldsymbol T_{r}(0)=[\boldsymbol I_{r\times r}^{T}~
 {\boldsymbol 0_{(M-r) \times
r}^{T}}]^T$\\
\bfseries {Update for each time instant} $i=1,\ldots,N$\\
~~~~~~$\bar{\boldsymbol x}(i)=\boldsymbol T_{r}^{H}(i-1)\boldsymbol x(i)$\\
~~~~~~$\bar{\boldsymbol a}(\theta_{n})=\boldsymbol
T_{r}^{H}(i-1)\boldsymbol a(\theta_{n})$\\
~~~~~~$\hat{{\boldsymbol R}}(i)=\alpha\hat{{\boldsymbol
R}}(i-1)+{\boldsymbol x}(i){\boldsymbol x}^{H}(i)$\\
~~~~~~$\hat{\bar{\boldsymbol R}}(i)=\alpha\hat{\bar{\boldsymbol
R}}(i-1)+\bar{\boldsymbol x}(i)\bar{\boldsymbol x}^{H}(i)$\\
~~~~~~$\bar{\boldsymbol g_{\theta}}(i)={\hat{\bar{\boldsymbol
R}}^{-1}(i)\bar{\boldsymbol a}(\theta_n)}/\big({\bar{\boldsymbol
a}^{H}(\theta_n)\hat{\bar{\boldsymbol R}}^{-1}(i)\bar{\boldsymbol
a}(\theta_n)}\big)$\\
~~~~~~$\boldsymbol T_r(i)=\frac{\hat{\boldsymbol
R}^{-1}(i)\boldsymbol a(\theta_n)}{\boldsymbol
a^H(\theta_n)\hat{\boldsymbol R}^{-1}(i)\boldsymbol
a(\theta_n)}\frac{\bar{\boldsymbol
g_{\theta}}^H(i)}{\|\bar{\boldsymbol g_{\theta}}(i)\|^2}$\\
\bfseries {Output power}\\
~~~~~~$P(\theta_{n})=1/\big(\bar{\boldsymbol
a}^{H}(\theta_{n})\hat{\bar{\boldsymbol R}}^{-1}\bar{\boldsymbol
a}(\theta_{n})\big)$\\
\bfseries {Polynomial rooting (optional)}\\
~~~~~~$Q(\theta)=\bar{\boldsymbol a}^H(\theta)\bar{\boldsymbol
R}^{-1}\bar{\boldsymbol a}(\theta)$\\
\hline
    \end{tabular}
    \end{small}
\end{table}

\subsection{Proposed JIO Algorithm with FBA}
The FBA technique is helpful to increase the resolution for DOA
estimation when the sources are correlated. It is based on the
averaging of the covariance matrix of identical overlapping arrays
and so requires an array of identical elements equipped with some
form of periodic structure, such as the ULA. For its application, we
split a ULA antenna array into a set of forward and conjugate
backward subarrays. 
 {The FBA preprocessing operates on $\boldsymbol x(i)$
to estimate the forward and backward subarray covariance matrices
that are averaged to get the forward/conjugate backward smoothed
covariance matrix. The JIO algorithm incorporated with the FBA
technique is termed JIO(FBA). }

In this work, we employ an efficient way to estimate the
forward/conjugate backward covariance matrix (see Eq. (3.22) in
\cite{Haardt4}). The resulting JIO(FBA) algorithm is summarized in
Table \ref{tab:JIO-FBA}, where $\boldsymbol\Pi_{M}\in\mathbb
C^{M\times M}$ is a matrix with ones on its antidiagonal and zeros
elsewhere, $\hat{\boldsymbol R}_{\textrm{fb}}(i)$ and
$\hat{\bar{\boldsymbol R}}_{\textrm{fb}}(i)$ are the full-rank and
the reduced-rank forward/backward averaged covariance matrices,
respectively. Note that $\hat{\boldsymbol R}(i)$ here is calculated
by using a time-averaged estimate, i.e., $\hat{{\boldsymbol
R}}(i)=\alpha\hat{{\boldsymbol R}}(i-1)+{\boldsymbol
x}(i){\boldsymbol x}^{H}(i)$. The proposed JIO(FBA) algorithm
employs the averaged $\hat{\boldsymbol R}_{\textrm{fb}}(i)$ and
$\hat{\bar{\boldsymbol R}}_{\textrm{fb}}(i)$ to compute
$\bar{\boldsymbol g}_{\theta,\textrm{fb}}(i)$ and $\boldsymbol
T_{\textrm{fb}}(i)$ for the output power with respect to each
scanning angle $\theta_n$. The computational complexity can be
significantly reduced by using a real-valued implementation
\cite{Haardt4}.

\begin{table}[t]
\centering
    \caption{The proposed JIO(FBA) algorithm}
    \label{tab:JIO-FBA}
    \begin{small}
    \begin{tabular}{|l|}
\hline
\bfseries {Initialization:}\\
~~~~~~$\boldsymbol T_{\textrm{fb}}(0)=[\boldsymbol I_{r\times
r}^{T}~  {\boldsymbol 0_{(M-r) \times
r }^{T}}]^T$\\
\bfseries {Update for each time instant} $i=1,\ldots,N$\\
~~~~~~$\bar{\boldsymbol a}_{\textrm{fb}}(\theta_n)=\boldsymbol
T_{\textrm{fb}}^H(i-1)\boldsymbol a(\theta_n)$\\
~~~~~~ {$\hat{{\boldsymbol R}}(i)=\alpha\hat{{\boldsymbol
R}}(i-1)+{\boldsymbol x}(i){\boldsymbol x}^{H}(i)$}\\
~~~~~~$\hat{\boldsymbol R}_{\textrm{fb}}(i)=\frac{1}{2}\big[\hat{\boldsymbol R}(i)+\boldsymbol \Pi_M\hat{\boldsymbol R}^{\ast}(i)\boldsymbol \Pi_M\big]$\\
~~~~~~$\hat{\bar{\boldsymbol
R}}_{\textrm{fb}}(i)=\boldsymbol T_{\textrm{fb}}^H(i-1)\hat{\boldsymbol R}_{\textrm{fb}}(i)\boldsymbol T_{\textrm{fb}}(i-1)$\\
~~~~~~$\bar{\boldsymbol
g}_{{\theta},\textrm{fb}}(i)={\hat{\bar{\boldsymbol
R}}_{\textrm{fb}}^{-1}(i)\bar{\boldsymbol
a}_{\textrm{fb}}(\theta_n)}/\big({\bar{\boldsymbol
a}_{\textrm{fb}}^{H}(\theta_n)\hat{\bar{\boldsymbol
R}}_{\textrm{fb}}^{-1}(i)\bar{\boldsymbol
a}_{\textrm{fb}}(\theta_n)}\big)$\\
~~~~~~$\boldsymbol T_{\textrm{fb}}(i)=\frac{\hat{\boldsymbol
R}_{\textrm{fb}}^{-1}(i)\boldsymbol a(\theta_n)}{\boldsymbol
a^H(\theta_n)\hat{\boldsymbol R}_{\textrm{fb}}^{-1}(i)\boldsymbol
a(\theta_n)}\frac{\bar{\boldsymbol
g}_{{\theta},{\textrm{fb}}}^H(i)}{\|\bar{\boldsymbol g}_{{\theta},{\textrm{fb}}}(i)\|^2}$\\
\bfseries {Output power}\\
~~~~~~$P(\theta_{n})=1/\big(\bar{\boldsymbol
a}_{\textrm{fb}}^{H}(\theta_{n})\hat{\bar{\boldsymbol
R}}_{\textrm{fb}}^{-1}\bar{\boldsymbol
a}_{\textrm{fb}}(\theta_{n})\big)$\\
\bfseries {Polynomial rooting (optional)}\\
~~~~~~$Q(\theta)=\bar{\boldsymbol a}^H(\theta)\bar{\boldsymbol
R}^{-1}\bar{\boldsymbol a}(\theta)$\\
\hline
    \end{tabular}
    \end{small}
\end{table}

\subsection{Proposed JIO-RLS Algorithm}

We utilize the matrix inversion lemma \cite{Haykin} to develop a
JIO-based RLS algorithm (JIO-RLS) for DOA estimation without the
matrix inverse. Specifically, defining
$\hat{\boldsymbol\Phi}(i)=\hat{\boldsymbol R}^{-1}(i)$, yields
\begin{equation}\label{22}
\boldsymbol
k(i)=\frac{\alpha^{-1}\hat{\boldsymbol\Phi}(i-1)\boldsymbol
x(i)}{1+\alpha^{-1}\boldsymbol
x^H(i)\hat{\boldsymbol\Phi}(i-1)\boldsymbol x(i)},
\end{equation}
\begin{equation}\label{23}
\hat{\boldsymbol\Phi}(i)=\alpha^{-1}\hat{\boldsymbol\Phi}(i-1)-\alpha^{-1}\boldsymbol
k(i)\boldsymbol x^H(i)\hat{\boldsymbol\Phi}(i-1),
\end{equation}
where $\boldsymbol k(i)\in\mathbb C^{M\times 1}$ is the Kalman gain
vector and $\hat{\boldsymbol\Phi}(0)=\delta\boldsymbol I_{M\times
M}$ with $\delta$ being a positive value that needs to be set for
numerical stability.

Given $\hat{\bar{\boldsymbol\Phi}}(i)=\hat{\bar{\boldsymbol
R}}^{-1}(i)$, we have
\begin{equation}\label{24}
\bar{\boldsymbol
k}(i)=\frac{\alpha^{-1}\hat{\bar{\boldsymbol\Phi}}(i-1)\bar{\boldsymbol
x}(i)}{1+\alpha^{-1}\bar{\boldsymbol
x}^H(i)\hat{\bar{\boldsymbol\Phi}}(i-1)\bar{\boldsymbol x}(i)},
\end{equation}
\begin{equation}\label{25}
\hat{\bar{\boldsymbol\Phi}}(i)=\alpha^{-1}\hat{\bar{\boldsymbol\Phi}}(i-1)-\alpha^{-1}\bar{\boldsymbol
k}(i)\bar{\boldsymbol x}^H(i)\hat{\bar{\boldsymbol\Phi}}(i-1),
\end{equation}
where $\bar{\boldsymbol k}(i)\in\mathbb C^{r\times1}$ is the
reduced-rank gain vector and
$\hat{\bar{\boldsymbol\Phi}}(0)=\bar{\delta}\boldsymbol I_{r\times
r}$ with  {$\bar{\delta}>0$}.

Substituting the recursive procedures (\ref{22})-(\ref{25}) into the
proposed JIO algorithm instead of the matrix inverse results in the
JIO-RLS algorithm, which is concluded in Table \ref{tab:JIO-RLS},
where $\delta$ and $\bar{\delta}$ are selected according to the
input signal-to-noise ratio (SNR) \cite{Haykin}, and
$\hat{\boldsymbol\Phi}$ is the estimate of the inverse of the
received covariance matrix after $N$ snapshots. The specific values
will be given in the simulations. The JIO-RLS algorithm retains the
positive feature of the iterative exchange of information between
the rank-reduction matrix and auxiliary reduced-rank vector, which
avoids the degradation of the resolution, and utilizes a recursive
procedure to compute $\hat{\boldsymbol R}^{-1}$ and
$\hat{\bar{\boldsymbol R}}^{-1}$ for the reduced complexity.
\begin{table}[t]
\centering
    \caption{Proposed JIO-RLS algorithm}
    \label{tab:JIO-RLS}
    \begin{small}
    \begin{tabular}{|l|}
\hline
\bfseries {Initialization:}\\
~~~~~$\boldsymbol T_{r}(0)=[\boldsymbol I_{r}^{T}~
 {\boldsymbol 0_{(M-r) \times
r}^{T}}]^T$;~~~$\delta,~\bar{\delta}=$\textrm{positive
constants};\\
~~~~~$\hat{\boldsymbol\Phi}(0)=\delta\boldsymbol I_{M\times M}$;~~~$\hat{\bar{\boldsymbol\Phi}}(0)=\bar{\delta}\boldsymbol I_{r\times r}$.\\
\bfseries {Update for each time instant} $i=1,\ldots,N$\\
~~~~~$\bar{\boldsymbol x}(i)=\boldsymbol T_{r}^{H}(i-1)\boldsymbol x(i)$\\
~~~~~$\bar{\boldsymbol a}(\theta_{n})=\boldsymbol
T_{r}^{H}(i-1)\boldsymbol a(\theta_{n})$\\
~~~~~$\bar{\boldsymbol
k}(i)=\frac{\alpha^{-1}\hat{\bar{\boldsymbol\Phi}}(i-1)\bar{\boldsymbol
x}(i)}{1+\alpha^{-1}\bar{\boldsymbol
x}^H(i)\hat{\bar{\boldsymbol\Phi}}(i-1)\bar{\boldsymbol x}(i)}$\\
~~~~~$\hat{\bar{\boldsymbol\Phi}}(i)=\alpha^{-1}\hat{\bar{\boldsymbol\Phi}}(i-1)-\alpha^{-1}\bar{\boldsymbol
k}(i)\bar{\boldsymbol x}^H(i)\hat{\bar{\boldsymbol\Phi}}(i-1)$\\
~~~~~$\bar{\boldsymbol g}_{\theta}(i)=\frac{\hat{\bar{\boldsymbol\Phi}}(i)\bar{\boldsymbol a}(\theta_{n})}{\bar{\boldsymbol a}^H(\theta_{n})\hat{\bar{\boldsymbol\Phi}}(i)\bar{\boldsymbol a}(\theta_{n})}$\\
~~~~~$k(i)=\frac{\alpha^{-1}\hat{\boldsymbol\Phi}(i-1)\boldsymbol
x(i)}{1+\alpha^{-1}\boldsymbol
x^H(i)\hat{\boldsymbol\Phi}(i-1)\boldsymbol x(i)}$\\
~~~~~$\hat{\boldsymbol\Phi}(i)=\alpha^{-1}\hat{\boldsymbol\Phi}(i-1)-\alpha^{-1}\boldsymbol
k(i)\boldsymbol x^H(i)\hat{\boldsymbol\Phi}(i-1)$\\
~~~~~$\boldsymbol T_r(i)=\frac{\hat{\boldsymbol\Phi}(i)\boldsymbol a(\theta_{n})}{\boldsymbol a(\theta_{n})^H\hat{\boldsymbol\Phi}(i)\boldsymbol a(\theta_{n})}\frac{\bar{\boldsymbol g}_{\theta}^H(i)}{\|\bar{\boldsymbol g}_{\theta}(i)\|^2}$\\
\bfseries {Output power}\\
~~~~~$P(\theta_{n})=1/\big(\bar{\boldsymbol
a}^{H}(\theta_{n})\hat{\bar{\boldsymbol\Phi}}\bar{\boldsymbol
a}(\theta_{n})\big)$\\
\bfseries {Polynomial rooting (optional)}\\
~~~~~~$Q(\theta)=\bar{\boldsymbol a}^H(\theta)\bar{\boldsymbol
R}^{-1}\bar{\boldsymbol a}(\theta)$\\
\hline
    \end{tabular}
    \end{small}
\end{table}

We can also use the FBA technique in the proposed JIO-RLS algorithm
to improve the resolution when the sources are correlated. The
full-rank and the reduced-rank forward/conjugate backward smoothed
covariance matrix can be estimated by $\hat{\boldsymbol
R}_{\textrm{fb}}(i)$ and $\hat{\bar{\boldsymbol
R}}_{\textrm{fb}}(i)$ in Table \ref{tab:JIO-FBA}. Using the matrix
inversion lemma for the calculation of
$\hat{\boldsymbol\Phi}_{\textrm{fb}}(i)=\hat{\boldsymbol
R}_{\textrm{fb}}^{-1}(i)$ and
$\hat{\bar{\boldsymbol\Phi}}_{\textrm{fb}}(i)=\hat{\bar{\boldsymbol
R}}_{\textrm{fb}}^{-1}(i)$, we have
\begin{equation}\nonumber
\begin{split}
&\hat{\boldsymbol\Phi}_{\textrm{fb}}(i)=2[\hat{\boldsymbol
\Phi}(i)-\frac{\hat{\boldsymbol\Phi}(i)\boldsymbol\Pi_M\Pi_M^H\hat{\boldsymbol\Phi}(i)}{\hat{\boldsymbol
\Phi}^{\ast}(i)+\boldsymbol\Pi_M^H\hat{\boldsymbol\Phi}(i)\boldsymbol\Pi_M}]\\
&\hat{\bar{\boldsymbol\Phi}}_{\textrm{fb}}(i)=2[\hat{\bar{\boldsymbol
\Phi}}(i)-\frac{\hat{\bar{\boldsymbol\Phi}}(i)\boldsymbol
T_{\textrm{fb}}^H(i-1)\boldsymbol\Pi_M\Pi_M^H\boldsymbol
T_{\textrm{fb}}(i-1)\hat{\bar{\boldsymbol\Phi}}(i)}{\hat{\bar{\boldsymbol
\Phi}}^{\ast}(i)+\boldsymbol\Pi_M^H\boldsymbol
T_{\textrm{fb}}(i-1)\hat{\bar{\boldsymbol\Phi}}(i)\boldsymbol
T_{\textrm{fb}}^H(i-1)\boldsymbol\Pi_M}],
\end{split}
\end{equation}
where both $\hat{\boldsymbol \Phi}(i)$ and $\hat{\bar{\boldsymbol
\Phi}}(i)$ are calculated by their recursive expressions, which have
been given in Table \ref{tab:JIO-RLS}. By using
$\hat{\boldsymbol\Phi}_{\textrm{fb}}(i)$ and
$\hat{\bar{\boldsymbol\Phi}}_{\textrm{fb}}(i)$ to replace
$\hat{\boldsymbol\Phi}(i)$ and $\hat{\bar{\boldsymbol\Phi}}(i)$ in
Table \ref{tab:JIO-RLS}, respectively, for calculating the
rank-reduction matrix and the auxiliary reduced-rank parameter
vector, we can compute the output power with respect to each
scanning angle and find the DOAs corresponding to the sources.

We have so far detailed the proposed reduced-rank scheme and the
derivations of the proposed algorithms. There are two points that
need to be interpreted here. First, the polynomial rooting technique
derived in Section III-B can be employed in the proposed algorithms
as a preprocessing step to save the computational cost. Second, the
proposed JIO algorithms could work without the exact knowledge of
the number of sources $q$. They jointly update the rank-reduction
matrix and auxiliary reduced-rank vector to calculate the output
power and scan possible angles to plot the output power spectrum,
which do not require information about $q$. Most existing algorithms
need this information since, for the subspace-based algorithms, $q$
is critical to construct the signal subspace, and, for the ML
algorithm, it is important to improve the resolution. However, $q$
has to be estimated using other techniques with more complex
procedures, which increase the computational cost.

\subsection{Model-Order Selection}

 {The selection of the rank $r$ is important to the
proposed algorithms since it determines how much information could
be retained in the reduced-rank received vector and thus impacts the
resolution. However, it does not mean that a larger $r$ always leads
to a better resolution. A large $r$ (e.g., close to $M$) increases
the dimension of the reduced-rank received vector, which may cause
redundancy and significantly increase the computational complexity.
On the other hand, a small $r$ saves the cost but may lose
information that is useful to improve the resolution. In most of the
existing subspace algorithms, $r$ equals $q$, which is employed in
the eigen-decomposition for the construction of the signal subspace.
In the proposed algorithms, $r$ could be set to be some specific
values that do not necessarily equal $q$. The range of values has
been obtained by experiments and the theoretical explanation is that
the subspace fitting performed by the JIO algorithms does not
require many bases in the subspace to result in a good performance.
This has been verified in the simulations.}

We introduce an adaptive approach for selecting the rank. We
describe a model-order selection method based on the MV criterion
computed by the rank-reduction matrix $\boldsymbol T_r^{(r)}$ and
the auxiliary reduced-rank parameter vector $\bar{\boldsymbol
g}^{(r)}$, which is
\begin{equation}\label{26}
\begin{split}
&\boldsymbol T_r^{(r)}=\begin{bmatrix}
 t_{1,1} & t_{1,2} & \ldots & t_{1,r_{\textrm{min}}} & \ldots &
 t_{1,r_{\textrm{max}}}\\
 t_{2,1} & t_{2,2} & \ldots & t_{2,r_{\textrm{min}}} & \ldots &
 t_{2,r_{\textrm{max}}}\\
 \vdots & \vdots & \vdots & \vdots & \vdots & \vdots\\
 t_{M,1} & t_{M,2} & \ldots & t_{M,r_{\textrm{min}}} & \ldots &
 t_{M,r_{\textrm{max}}}\\ \end{bmatrix},\\
&\bar{\boldsymbol g}_{\theta}^{(r)}=\begin{bmatrix}
 \bar{g}_{1} & \bar{g}_{2} & \ldots & \bar{g}_{r_{\textrm{min}}} & \ldots &
 \bar{g}_{r_{\textrm{max}}}\\ \end{bmatrix}^{T},
\end{split}
\end{equation}
where the superscript $(\cdot)^{(r)}$ denotes the rank used for the
adaptation at each time instant, $r_{\textrm{min}}$ and
$r_{\textrm{max}}$ are the minimum and maximum ranks allowed,
respectively.

 {The rank is adopted automatically based on the
exponentially-weighted a $\textit{posteriori}$ MV criterion used to
derive the rank-reduction matrix and the auxiliary reduced-rank
vector, which is}
\begin{equation}\label{27}
\begin{split}
&\mathcal {J}_{\textrm{PLS}}\big(\boldsymbol T_r^{(r)}(i-1),
\bar{\boldsymbol
g}^{(r)}(i-1)\big)=\\
&~~~~~~~~~~\sum_{l=1}^i\varrho^{i-l}|\bar{\boldsymbol
g}_{\theta}^{(r)H}(i-1)\boldsymbol T_r^{(r)H}(i-1)\boldsymbol
x(l)|^2,
\end{split}
\end{equation}
where $\varrho$ is the exponential weight factor that is required as
the optimal rank $r$ can change as a function of the time instant
$i$. For each time instant, $\boldsymbol T_r^{(r)}(i)$ and
$\bar{\boldsymbol g}_{\theta}^{(r)}(i)$ are computed for a selected
$r$ according to \eqref{27}. The developed model-order selection
method is given by
\begin{equation}\label{28}
r_{\textrm{opt}}=\textrm{arg}\min_{r_{\textrm{min}}\leq r\leq
r_{\textrm{max}}}\mathcal{J}_{\textrm{PLS}}\big(\boldsymbol
T_r^{(r)}(i-1), \bar{\boldsymbol g}_{\theta}^{(r)}(i-1)\big),
\end{equation}
where $r$ is an integer ranging between $r_{\textrm{min}}$ and
$r_{\textrm{max}}$. For each $\theta_n$, we calculate $\boldsymbol
T_r^{(r)}$ and $\bar{\boldsymbol g}_{\theta}^{(r)}$ until
$r=r_{\textrm{max}}$, and then scan $(r)$ from $r_{\textrm{min}}$ to
$r_{\textrm{max}}$ to find a pair of $\{\boldsymbol T_r^{(r)},
\bar{\boldsymbol g}_{\theta}^{(r)}\}$ that satisfy \eqref{28}.
 {It should be noted that the model-order selection
algorithm is not very expensive because it re-uses entries of the
rank-reduction matrix which is initially computed with $r_{\rm max}$
and the matrix inverse (no extra matrix inverse is required). The
algorithm then computes extra terms for the optimization in
\eqref{28}.  The corresponding $(r)$ is the most appropriate value
with respect to the current time instant. We found that the range
for which the rank $r$ has an impact on the resolution is very
limited, being from $r_{\textrm{min}}=3$ to $r_{\textrm{max}}=7$.
These values are rather insensitive to the number of users in the
system, to the number of sensor elements, and work effectively for
the studied scenarios.  {If $q$ is larger than $r_max$ then there is
a potential risk of undermodelling. However, the method proposed
concentrates the energy of the subspace in a different way as
compared to an eigen-based technique. This is the reason why it was
found that $r_max=7$ was enough for the scenarios studied.}
Alternatively, an additional mechanism can be used to adjust
$r_{\textrm{min}}$ and $r_{\textrm{max}}$. The model-order selection
procedure involves additional complexity corresponding to the
computation of the cost function in \eqref{28} and requires
$3(r_{\rm max} -r_{\rm min}) +1$ additions and a sorting algorithm
to find the best model order according to \eqref{28}. It is
efficient to combine this approach with the polynomial rooting
technique to select the most appropriate rank for the proposed
algorithms.}

\subsection{Complexity Analysis}
Considering the computational cost, Capon's method, MUSIC, and
ESPRIT work with $O(M^3)$ due to the matrix inverse and the
eigen-decomposition, respectively. The recent AV and CG algorithms
pay a higher cost \cite{Semira} due to the generation of the signal
subspace. The API subspace tracking approach estimates the signal
subspace of the MUSIC and the ESPRIT with lower complexity
$O(qM+q^3)$. However, its complexity becomes relatively high as the
number of sources $q$ becomes large. For the proposed algorithms,
the JIO algorithm requires $O(M^3+r^3)$ due to the matrix inverse.
The JIO-RLS algorithm requires $O(M^2+r^2)$ due to the use of the
matrix inversion lemma \cite{Haykin}. It is worth noting that the
cost of computing $\hat{\boldsymbol\Phi}$ is saved after the
procedure of the first scanning angle since it is invariable for the
rest of the search.

We provide a comparison of the computational complexity for the
proposed and existing algorithms in Table \ref{tab:complexity},
where $M$ denotes the number of sensor elements, $q$ is the number
of sources, $\Delta$ is the search step, $D$ is the iteration number
for the CG algorithm, and $r$ is the rank. Note that the cost of
$r^3$ (or $r^2$) is much less than that of $M^3$ (or $M^2$) since
$r$ is always much smaller than $M$ for sufficiently large arrays.
Our studies reveal that the range for which the rank $r$ has a
positive impact on the performance is limited among a set of small
values. This fact has been referred to the previous section and will
be verified in the simulations. The complexity of the algorithms
equipped via the FBA technique is not shown in this table since it
is viewed as a preprocessing step and requires nearly the same cost
for all the methods.

\begin{table}[htp]
\centering
    \caption{Comparison of The Computational Complexity}     
    \label{tab:complexity}
    \begin{footnotesize}
    \begin{tabular}{|l l l|}
\hline
\bfseries Algorithms & \bfseries Complexity & \bfseries Main Procedures\\
 \hline
Capon       &$O(M^3)$  &  Matrix inverse (grid search)  \\
MUSIC       &$O(M^3)$   &  Eigen-decomposition (grid search)\\
MUSIC(API)  &$O(qM+q^3)$ & Subspace tracking (grid search)\\
ESPRIT      & $O(M^3)$  & Eigen-decomposition\\
ESPRIT(API) & $O(qM+q^3)$ & Subspace tracking\\
AV          & $O((180/\Delta)qM^2)$ & Construction of signal
subspace (grid search)\\
CG          & $O((180/\Delta)DM^2)$ & Construction of signal
subspace (grid search)\\
JIO         & $O(M^3+(180/\Delta)r^3)$ & Matrix inverse and reduced-rank process (grid search)\\
JIO-RLS     & $O(M^2+(180/\Delta)r^2)$ & Matrix inversion lemma
and  reduced-rank processing (grid search)\\
Root JIO-RLS     & $O(M^2+r^2)$ & Matrix inversion lemma and
reduced-rank processing (polynomial rooting)\\
 \hline
    \end{tabular}
    \end{footnotesize}
\end{table}

\section{Simulations}

In this section, we evaluate the probability of resolution of the
proposed JIO algorithms and compare them with a number of existing
DOA estimation algorithms. The probability of resolution for two
sources is defined as  { $P_r[|\hat{\theta}_1-\theta_1| {\rm and}
|\hat{\theta}_2 - \theta_2|<|\theta_1-\theta_{2}|/2]$}
\cite{Grover}, \cite{Stoica}. We compare the proposed JIO algorithms
with Capon's method, the MUSIC and ESPRIT subspace-based methods
with and without the API subspace tracking implementation, the
projected companion matrix MUSIC (PCM-MUSIC) \cite{korso}, the
 {fast root MUSIC \cite{ren}} and the ML method. In
all simulations, binary phase shift keying (BPSK) sources separated
by $3^o$ with powers $\sigma_s^2=1$ are considered and the noise is
spatially and temporally white Gaussian. All the results are
averaged over $1000$ runs. The search step is $\Delta^o=0.5^o$.
 {The forgetting factor $\alpha$ corresponds to a
coherence window of $1/(1-\alpha)$ snapshots. When the number of
snapshots is small which is the case of interest in this work, then
there is no significant impact of using an $\alpha$ different from
$1$ but as the number of snapshots is increased the forgetting
factor should match the coherence window of the time-varying process
in order to track dynamic sources. The forgetting factor is
important in non-stationary scenarios in which there is need to
discard past data to obtain more accurate estimates. } {The diagonal
loading (or regularization) has been used for all the methods and
the parameters have been optimized for each method in order to
ensure a fair comparison.} Simulations are performed for a ULA with
half a wavelength inter-element spacing for a generic application of
all studied algorithms.

\begin{figure}[htb]
\begin{minipage}[h]{1.0\linewidth}
  \centering
  \centerline{\epsfig{figure=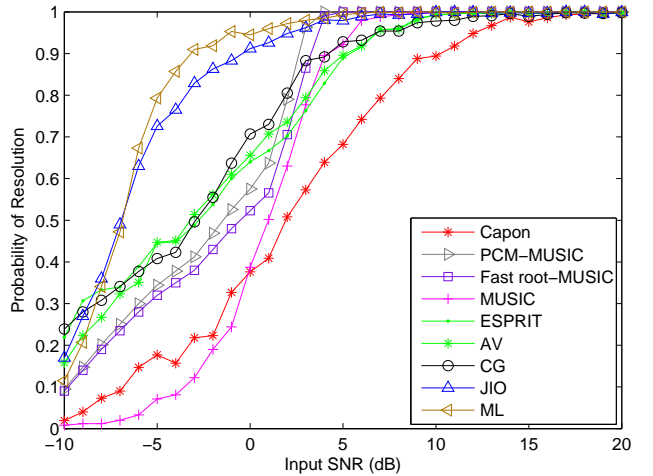,scale=0.65}} \vspace{-0.5em}\caption{Probability of resolution versus input
  SNR with $\alpha=0998$, $M=40$, $N=10$, $q=2$, $\tau=0.9$, $r_{\rm opt}=4$.}
\label{fig:cor}
\end{minipage}
\end{figure}

In Fig. \ref{fig:cor}, we assess the impact of the correlated
sources on the performance of the proposed and existing algorithms.
The array size is $M=40$ and the number of snapshots $N=10$ is
fixed. There are $q=2$ highly correlated sources in the system with
correlation value $\tau=0.9$, which are generated as follows
\cite{Grover}:
$$s_1\sim\mathcal {N}(0, \sigma_s^2)~~\textrm{and}~~s_2=\tau
s_1+\sqrt{1-\tau^2}s_3,$$ where $s_3\sim\mathcal {N}(0,
\sigma_s^2)$. The rank for the proposed JIO algorithm is $r=4$. We
have verified the rank among $r\in[1,8]$ and found that $r=4$ is the
most appropriate value. Although the values that are higher than
$r=4$ could also achieve a relatively high probability of
resolution, they result in a higher computational load if
$r\rightarrow M$. The probability of resolution is plotted against
the input SNR values. From Fig. \ref{fig:cor}, we can see that the
ML algorithm is superior to the other analyzed algorithms. However,
it requires a higher computational cost than the remaining
techniques. The proposed JIO algorithm outperforms other existing
algorithms for different SNR values.

\begin{figure}[htb]
\begin{minipage}[h]{1.0\linewidth}
  \centering
  \centerline{\epsfig{figure=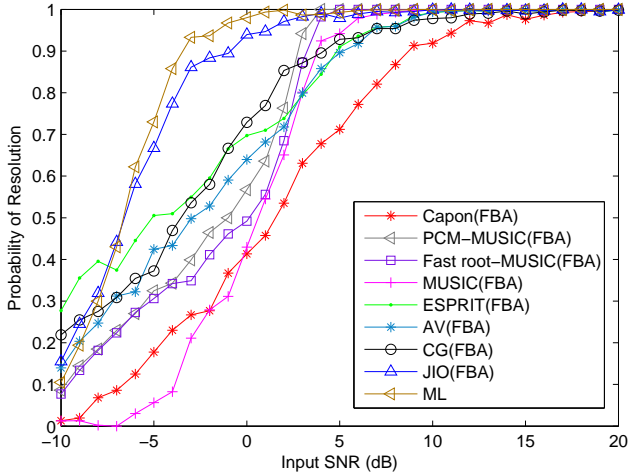,scale=0.65}} \vspace{-0.5em}\caption{Probability of resolution versus input
  SNR with $\alpha=0998$, $M=40$, $N=10$, $q=2$, $\tau=0.9$, $r_{\rm opt}=4$.}
\label{fig:cor_fba}
\end{minipage}
\end{figure}

 {In Fig. \ref{fig:cor_fba}, we consider the same
scenario as in Fig. \ref{fig:cor} and show the performance of the
proposed and existing algorithms equipped with the FBA technique.
The ML algorithm is included in this experiment to provide a
comparison with Fig. \ref{fig:cor}. It is clear that the FBA
technique is useful to the studied algorithms for dealing with the
problem of the highly correlated sources. The probability of
resolution for each algorithm is improved under this case in
comparison with its conventional counterpart  { shown in Fig.
\ref{fig:cor}}. The proposed JIO algorithm shows a high performance
that is close to the ML and superior to the others.}

\begin{figure}[htb]
\begin{minipage}[h]{1.0\linewidth}
  \centering
  \centerline{\epsfig{figure=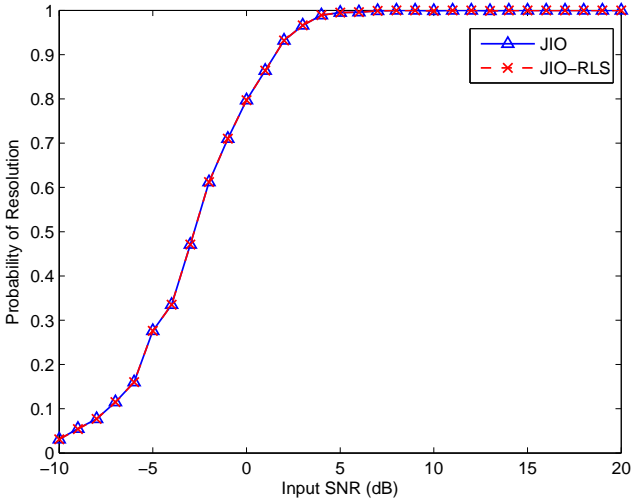,scale=0.65}} \vspace{-0.5em}\caption{Probability of resolution versus input
  SNR with $M=40$, $N=20$, $q=10$, $\alpha=0.998$, $\delta=\bar{\delta}=1\times10^{-3}$, $r_{\rm opt}=5$.}
\label{fig:jio_compare}
\end{minipage}
\end{figure}

Before further experiments, we evaluate the performance of the
proposed JIO and JIO-RLS algorithms, which is shown in Fig.
\ref{fig:jio_compare}. In this experiment, we set the sources to be
uncorrelated but increase the number of sources by setting $q=10$.
The number of snapshots is $N=20$ and the array size is $M=40$. From
Fig. \ref{fig:jio_compare}, we find that the proposed JIO and
JIO-RLS algorithms show nearly the same probability of resolution
with respect to different input SNR values. The same behavior is
observed for correlated sources with the use of the FBA technique.
These results show that the proposed JIO-RLS algorithm is an
efficient alternative to implement the JIO method. In what follows,
we will focus on the JIO-RLS algorithm and its comparison to other
techniques.

\begin{figure}[htb]
\begin{minipage}[h]{1.0\linewidth}
  \centering
  \centerline{\epsfig{figure=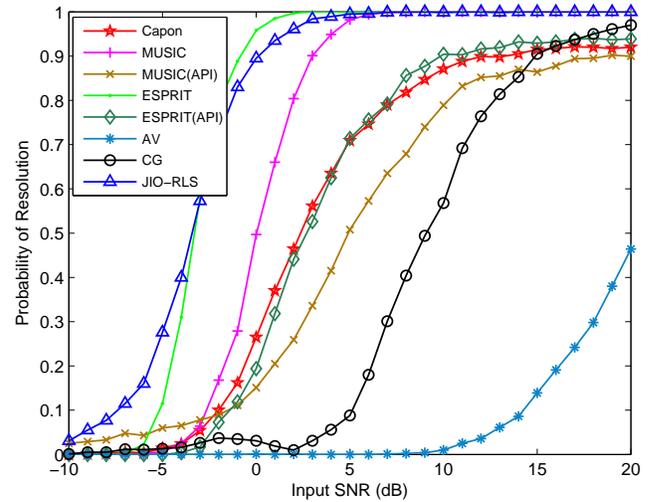,scale=0.65}} \vspace{-0.5em}\caption{Probability of resolution versus input
  SNR with $M=40$, $N=20$, $q=10$, $\alpha=1$, $\delta=\bar{\delta}=1\times10^{-3}$, $r_{\rm opt}=5$.}
\label{fig:more}
\end{minipage}
\end{figure}

 {In the next two experiments, the scenario is the
same as in Fig.\ref{fig:jio_compare}. We evaluate the probability of
resolution and the root mean-square error (RMSE) performance of the
proposed JIO-RLS algorithm. Note that the RMSE is computed by
averaging over the number of sources in the scenario. Furthermore,
we employ the polynomial rooting provided in Section III-B to reduce
the search length for the JIO-RLS, which further reduces the
complexity. In Fig. \ref{fig:more}, the curves between the proposed
and ESPRIT algorithms intersect when the input SNR values increase.
The proposed algorithm exhibits its ability to work with low SNR
values. ESPRIT uses the eigen-decomposition for estimating the
signal subspace. ESPRIT with the API approach \big(ESPRIT(API)\big)
performs direction finding with a low-complexity implementation.
However, this performance is poor with a small number of snapshots
and thus results in a low probability of resolution, so does
MUSIC(API). The AV and CG algorithms also show a low performance
when many sources are present in the system.}

\begin{figure}[htb]
\begin{minipage}[h]{1.0\linewidth}
  \centering
  \centerline{\epsfig{figure=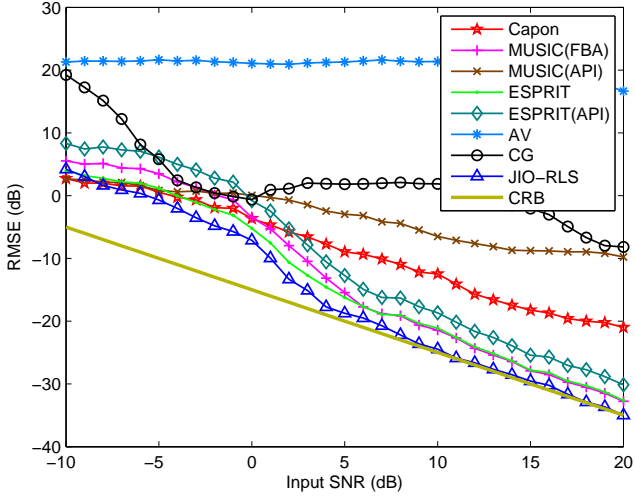,scale=0.65}} \vspace{-0.5em}\caption{RMSE versus input
  SNR with $M=40$, $N=20$, $q=10$, $\alpha=1$, $\delta=\bar{\delta}=1\times10^{-3}$, $r_{\rm opt}=5$.}
\label{fig:more_rmse}
\end{minipage}
\end{figure}

Fig. \ref{fig:more_rmse} presents the RMSE performance for the
proposed and existing algorithms under the same scenario as Fig.
\ref{fig:more} and compare them with the Cram\'{e}r-Rao bound (CRB).
The RMSE of the proposed JIO-RLS algorithm always keeps a lower
level than those of other existing algorithms with different input
SNR values. Its values are around $10$ dB higher than the CRB in the
threshold region and then approach the CRB curve with the increase
of the SNR. The eigen-decomposition algorithms (MUSIC and ESPRIT)
are superior to the AV and CG algorithms in this example.

\begin{figure}[htb]
\begin{minipage}[h]{1.0\linewidth}
  \centering
  \centerline{\epsfig{figure=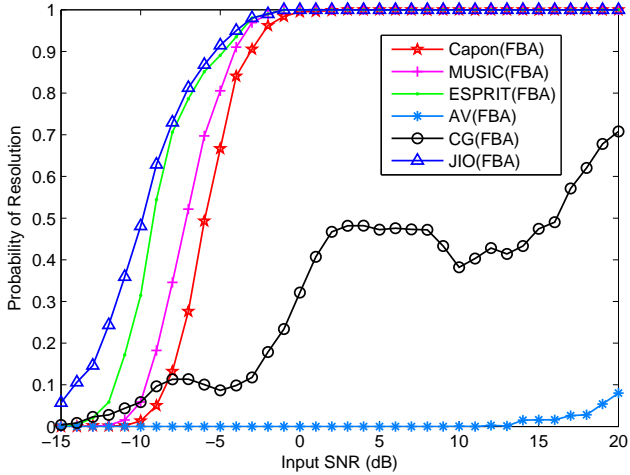,scale=0.65}} \vspace{-0.5em}\caption{Probability of resolution versus input
  SNR with $M=40$, $N=10$, $q=10$, $\alpha=1$, $\tau=0.9$, $\delta=\bar{\delta}=1\times10^{-3}$, $r_{\rm opt}=5$.}
\label{fig:more_fba}
\end{minipage}
\end{figure}

\begin{figure}[htb]
\begin{minipage}[h]{1.0\linewidth}
  \centering
  \centerline{\epsfig{figure=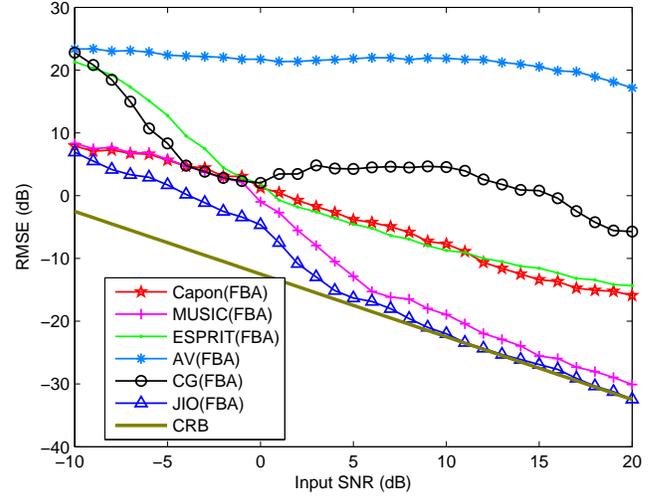,scale=0.65}} \vspace{-0.5em}\caption{RMSE versus input
  SNR with $M=40$, $N=10$, $q=10$, $\alpha=1$, $\tau=0.9$, $\delta=\bar{\delta}=1\times10^{-3}$, $r_{\rm opt}=5$.}
\label{fig:more_fba_rmse}
\end{minipage}
\end{figure}

 { Fig. \ref{fig:more_fba} and Fig.
\ref{fig:more_fba_rmse} examine the performance of the proposed and
existing algorithms with the FBA technique in a severe scenario,
where many sources ($q=10$) are present in the system and the number
of snapshots is low ($N=10$). {Two sources are highly correlated as
explained in the beginning of this section.} The number of sensor
elements in the array is $M=40$. From Fig. \ref{fig:more_fba}, the
AV with the FBA technique fails to resolve the DOA at most input SNR
values. The CG(FBA) algorithm provides a good resolution but is
unstable with respect to different SNR values. The Capon's, MUSIC,
ESPRIT, and the proposed algorithms exhibit relatively high
resolutions at high SNR values. The proposed JIO-RLS(FBA) algorithm
works well even with very low SNR values. Fig.
\ref{fig:more_fba_rmse} reflects the RMSE performance of the studied
algorithms under the same condition. The proposed JIO algorithm
approaches the CRB asymptotically and follows the trend of the CRB
as the SNR value increases to $5$ dB.}

\begin{figure}[htb]
\begin{minipage}[h]{1.0\linewidth}
  \centering
  \centerline{\epsfig{figure=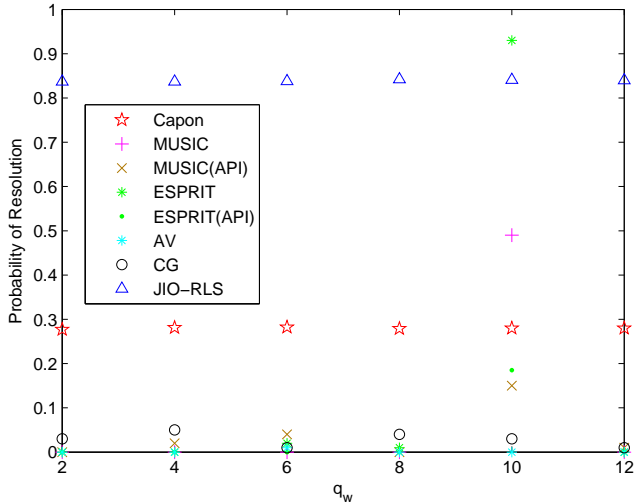,scale=0.65}}
  \vspace{-0.5em}\caption{Probability of resolution versus $q_w$ with
  $M=40$, $N=20$, SNR$=0$ dB, $q=10$, $\alpha=0.998$, $\delta=\bar{\delta}=1\times10^{-3}$, $r_{\rm opt}=5$.}
\label{fig:jio_rls_vs_qw_final}
\end{minipage}
\end{figure}

 {In the following results, we consider a situation
where the receiver antenna does not know exactly the information of
the number of sources $q$. This is more practical since the exact
$q$ has to be determined by procedures with extra computational cost
and time or by resorting noise threshold with subspace tracking
algorithms \cite{Badeau}. The purpose is assess the robustness of
the methods in the presence of errors in the model order by
measuring the performance degradation of such techniques. The
scenario is the same as in Fig. \ref{fig:more}. We set the input SNR
to $0$ dB and examine the probability of resolution of the proposed
and existing algorithms with respect to different values of $q_w$.
From Fig. \ref{fig:jio_rls_vs_qw_final}, the proposed and existing
algorithms are evaluated with a variable $q_w$ The results show that
the proposed JIO and Capon's algorithms are not significantly
affected by different values of $q_w$, whereas the performance of
the other studied algorithms is significantly degraded for $q_w\neq
q$. Note that, in this scenario, the number of snapshots $N=20$ is
quite small as compared to the number of sensors $M=40$, and this is
not sufficient for the existing subspace-based algorithms to
construct the signal subspace.}

\begin{figure}[htb]
\begin{minipage}[h]{1.0\linewidth}
  \centering
  \centerline{\epsfig{figure=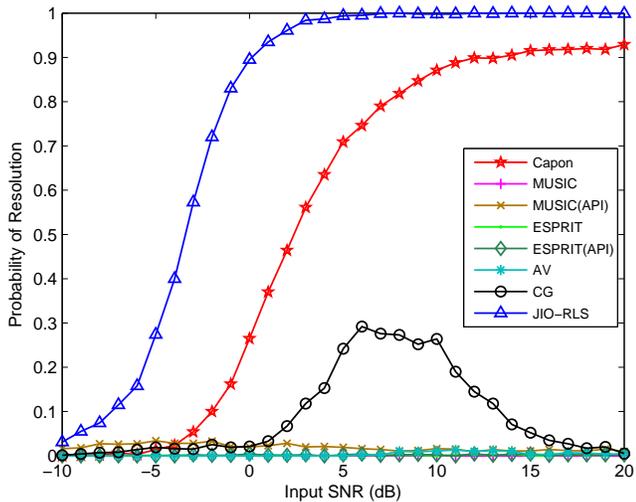,scale=0.65}}
  \vspace{-0.5em}\caption{Probability of resolution
  versus $q_w$ with $M=40$, $N=20$, $q=10$, $q_w=9$, $\alpha=0.998$, $\delta=\bar{\delta}=1\times10^{-3}$, $r_{\rm opt}=5$.}
\label{fig:wrongq}
\end{minipage}
\end{figure}

\begin{figure}[htb]
\begin{minipage}[h]{1.0\linewidth}
  \centering
  \centerline{\epsfig{figure=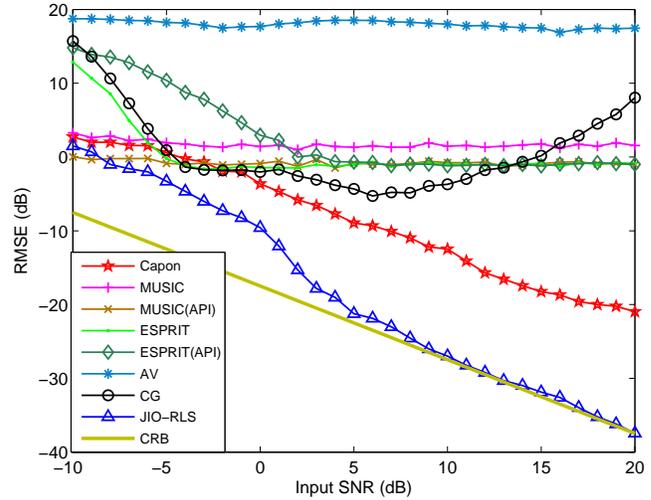,scale=0.65}} \vspace{-0.5em}\caption{RMSE versus input
  SNR versus $q_w$ with $M=40$, $N=20$, $q=10$, $q_w=9$, $\alpha=0.998$,
  $\delta=\bar{\delta}=1\times10^{-3}$, $r_{\rm opt}=5$.}
\label{fig:wrongq_rmse}
\end{minipage}
\end{figure}

In Fig. \ref{fig:wrongq} and Fig. \ref{fig:wrongq_rmse}, we set an
incorrect number of sources $q_{\textrm{w}}=9$ for the receiver and
show the performance versus different input SNR values. Fig.
\ref{fig:wrongq} exhibits the probability of resolution for the
proposed and existing algorithms. The eigen-decomposition and their
related API algorithms fail to solve the DOA estimation problem at
all input SNR values since $q$ is critical to the estimation of the
signal and noise subspaces. Also, the design of the AV basis and CG
residual vectors depends strongly on $q$ and cannot achieve a good
direction finding. Capon's method works well under this condition
since it is insensitive to the number of sources. The same result
holds for the proposed JIO-RLS algorithm, which outperforms Capon's
method since the joint optimization between the rank reduction
matrix and the auxiliary reduced-rank parameter vector leads to an
improved performance for the proposed scheme. We also provide the
RMSE performance in Fig. \ref{fig:wrongq_rmse}. The subspace-based
algorithms always keep a high RMSE level (above $0$ dB) and do not
approach the CRB. The proposed algorithm is not significantly
influenced by $q_\textrm{w}$ and retains the same trend as the CRB,
as depicted in Fig. \ref{fig:more}. We also consider the algorithms
with the FBA technique in this condition and obtain a comparable
result.

\section{Concluding Remarks}
We have introduced a novel reduced-rank scheme based on the joint
iterative optimization of a rank-reduction matrix and an auxiliary
parameter vector for DOA estimation. In the proposed scheme, the
dimension of the received vector is reduced by the rank-reduction
matrix, and the resulting vector is processed by the auxiliary
reduced-rank parameter vector for calculating the output power. It
provides an iterative exchange of information between the estimated
quantities and thus leads to an improved performance. The DOAs of
the sources are located by scanning the possible angles and plotting
the output power spectrum. The proposed JIO algorithms have been
implemented to iteratively estimate the rank-reduction matrix and
the auxiliary parameter vector according to the MV design criterion.
The polynomial rooting technique has been incorporated in the
proposed JIO algorithms to save some computational cost. We have
employed the FBA preprocessing to deal with the problem of the
highly correlated sources. The proposed algorithms also work well
without the exact information of the number of sources. Simulations
have shown that the proposed JIO algorithms achieve a superior
resolution over the existing algorithms in the scenarios where many
sources are present in the system, the array size is large, and the
number of snapshots is small.

\begin{appendix}
\section*{Derivation of the rank-reduction matrix}
Equating (\ref{7}) to a zero matrix and post multiplying the terms
by $\bar{\boldsymbol g}_{\theta}$ yields
\begin{equation}\label{a1}
\boldsymbol T_r\bar{\boldsymbol
g}_{\theta}=-\lambda_{T_r^{\ast}}\boldsymbol R^{-1}\boldsymbol
a(\theta).
\end{equation}

Given $\boldsymbol f=-\lambda_{T_r^{\ast}}\boldsymbol
R^{-1}\boldsymbol a(\theta)$, the matrix $\boldsymbol T_r$ can be
viewed as finding a solution to the linear equation $\boldsymbol
T_r\bar{\boldsymbol g}_{\theta}=\boldsymbol f$. Assuming
$\bar{\boldsymbol g}_{\theta}\neq\boldsymbol 0$, there exist
multiple $\boldsymbol T_r$ satisfying the linear equation in
general. Thus, we derive the minimum Frobenius-norm solution for
stability. Let us express the quantities involved by
\begin{equation}\label{a2}
\boldsymbol T_r=[\bar{\boldsymbol t}_1, \bar{\boldsymbol t}_2,
\ldots, \bar{\boldsymbol t}_{M}]^H;~~~~\boldsymbol f=[f_1, f_2,
\ldots, f_M]^T,
\end{equation}
where $\bar{\boldsymbol t}_{j}=[\bar{t}_{j,1}^{\ast}, \ldots,
\bar{t}_{j,r}^{\ast}]^T\in\mathbb C^{r\times1}$ with $j=1, \ldots,
M$.

The computation of the minimum Frobenius-norm solution transfers to
the following $M$ subproblems:
\begin{equation}\label{a3}
\textrm{min}\|\bar{\boldsymbol
t}_j\|^2,~~~\textrm{subject~to~}\bar{\boldsymbol
t}_{j}^H\bar{\boldsymbol g}_{\theta}=f_j.
\end{equation}

The solution to (\ref{a3}) is the projection of $\bar{\boldsymbol
t}_j$ onto the hyperplane $\mathcal{H}_j=\{\bar{\boldsymbol
t}_j\in\mathbb C^{r\times1}\}:\bar{\boldsymbol
t}_j^H\bar{\boldsymbol g}_{\theta}=f_j$, which is given by
\begin{equation}\label{a4}
\bar{\boldsymbol t}_j=f_j^{\ast}\frac{\bar{\boldsymbol
g}_{\theta}}{\|\bar{\boldsymbol g}_{\theta}\|^2}.
\end{equation}

Thus, the rank-reduction matrix can be expressed by
\begin{equation}\label{a5}
\boldsymbol T_r=\boldsymbol f\frac{\bar{\boldsymbol
g}_{\theta}^H}{\|\bar{\boldsymbol g}_{\theta}\|^2}.
\end{equation}

Substituting $\boldsymbol f=-\lambda_{T_r^{\ast}}\boldsymbol
R^{-1}\boldsymbol a(\theta)$ into (\ref{a5}) and following the
constraint in (\ref{5}), we get the $\lambda_{T_r^{\ast}}$, which is
\begin{equation}\label{a6}
\lambda_{T_r^{\ast}}=-\frac{1}{\boldsymbol a^H(\theta)\boldsymbol
R^{-1}\boldsymbol a(\theta)}.
\end{equation}

From $\boldsymbol f$, $\lambda_{T_r^{\ast}}$, and $\boldsymbol T_r$
in (\ref{a5}), we have the expression of the rank-reduction matrix
in (\ref{8}).

\section*{Derivation of the Reduced-Rank Vector $\bar{\boldsymbol g}_{\theta}$}
 {
Equating (\ref{9}) to zero, we have
\begin{equation}
\bar{\boldsymbol g}_{\theta}=
\lambda_{\bar{g}_{\theta}^{\ast}}{\bar{\boldsymbol
R}^{-1}\bar{\boldsymbol a}(\theta)}, \label{a7}
\end{equation}
where $\bar{\boldsymbol R}=\mathbb E[\bar{\boldsymbol
x}(i)\bar{\boldsymbol x}^H(i)]\in\mathbb C^{r\times r}$ is the
reduced-rank covariance matrix. Substituting \eqref{a7} into the
constraint and equating the terms, we obtain
\begin{equation}
\lambda_{\bar{g}_{\theta}^{\ast}} =  (\bar{\boldsymbol
a}^H(\theta)\bar{\boldsymbol R}^{-1} \bar{\boldsymbol
a}(\theta))^{-1}, \label{a8}
\end{equation}
Substituting $\lambda_{\bar{g}_{\theta}^{\ast}}$ into \eqref{a7}, we
obtain
\begin{equation}
\bar{\boldsymbol g}_{\theta}=\frac{\bar{\boldsymbol
R}^{-1}\bar{\boldsymbol a}(\theta)}{\bar{\boldsymbol
a}^H(\theta)\bar{\boldsymbol R}^{-1}\bar{\boldsymbol a}(\theta)},
\end{equation}
where $\bar{\boldsymbol a}(\theta)=\boldsymbol T_r^H\boldsymbol
a(\theta)\in\mathbb C^{r\times1}$ is the reduced-rank steering
vector with respect to the current scanning angle.}

\end{appendix}

\end{document}